\documentclass[aps,prl,twocolumn,longbibliography,floatfix,10pt,superscriptaddress]{revtex4-2}
\usepackage[utf8]{inputenc}
\usepackage{mathrsfs}
\usepackage{natbib}
\usepackage{graphicx}
\usepackage{latexsym}
\usepackage{amssymb}
\usepackage{amsmath}
\usepackage{amsfonts}
\usepackage{gensymb}
\usepackage{bm}
\usepackage{hyperref}
\usepackage{braket}
\usepackage{physics}
\usepackage{bbold}
\usepackage[caption=false]{subfig}
\usepackage[dvipsnames]{xcolor}
\usepackage[normalem]{ulem}
\usepackage{siunitx}
\usepackage{mhchem}
\usepackage{multirow}
\usepackage{soul}

\usepackage{slashed}

\definecolor{myforestgreen}{RGB}{34,139,34}

\newcommand{\PRLsec}[1]{\emph{#1---}}

\newcommand{\supplementarysection}{%
  \setcounter{figure}{0}
  \let\oldthefigure\thefigure
  \renewcommand{\thefigure}{S\oldthefigure}
  \setcounter{section}{0}
  \let\oldthesection\thesection
  \renewcommand{\thesection}{S\oldthesection}
  \setcounter{equation}{0}
  \let\oldtheequation\theequation
  \renewcommand{\theequation}{S\oldtheequation}
  \setcounter{table}{0}
  \let\oldthetable\thetable
  \renewcommand{\thetable}{S\oldthetable}
}

\newenvironment{dfn}{{\vspace*{1ex} \noindent \bf Definition }}{\vspace*{1ex}}

	\newcommand{\beq}{\begin{eqnarray}}
	\newcommand{\eeq}{\end{eqnarray}}
	\newcommand{\bea}{\begin{eqnarray}\begin{aligned}}
	\newcommand{\eea}{\end{aligned}\end{eqnarray}}

\begin{document}

\title{Ferrofluids under oscillatory magnetic fields}

\author{Taige Wang}
\affiliation{No.2 High School of East China Normal University, Shanghai 201203, PRC}
\affiliation{Department of Physics, Harvard University, Cambridge, MA 02138, USA \looseness=-2}
\affiliation{Department of Physics, Massachusetts Institute of Technology, Cambridge, MA 02139, USA \looseness=-2}

\author{Kaiyuan Gu}
\affiliation{No.2 High School of East China Normal University, Shanghai 201203, PRC}
\affiliation{Department of Physics, Princeton University, Princeton, New Jersey 08544, USA \looseness=-2}

\author{Anzhou Wang}
\affiliation{No.2 High School of East China Normal University, Shanghai 201203, PRC}
\affiliation{JILA, NIST and University of Colorado, Boulder, Colorado 80309, USA \looseness=-2}

\author{Zhentang Wang}
\affiliation{No.2 High School of East China Normal University, Shanghai 201203, PRC}

\begin{abstract}
Ferrofluids exhibit two canonical interfacial instabilities, a static Rosensweig (normal-field) instability that produces a lattice of peaks and a dynamical Faraday instability that produces parametrically excited standing waves. Here we present a systematic phase diagram of ferrofluid surface states driven by a purely AC vertical magnetic field with zero mean. Scanning a broad range of frequencies and field amplitudes, we resolve two robust branches: a Faraday-wave regime that includes a stable square lattice and a Rosensweig-like peak--valley regime indistinguishable in morphology from Rosensweig peaks. The Faraday-onset boundary is well described by a power law close to $\sqrt{f}$, while the Rosensweig-like peak onset becomes essentially frequency independent at low viscosity. The wave vector of the square lattice grows linearly with frequency over our accessible band. We present a surface-wave theory that captures the full phenomenology, including the emergence of Rosensweig peaks under zero-mean AC driving, the near-$\sqrt{f}$ scaling of the phase boundaries, the linear growth of the selected wave vector with frequency, and the preference for square over hexagonal lattices.
\end{abstract}

\maketitle

\PRLsec{Introduction}
Ferrofluids are colloidal suspensions of magnetic nanoparticles that retain the flow properties of a liquid while coupling strongly to an external magnetic field \cite{RosensweigBook,Rosensweig1987ARFM}. This coupling is especially transparent at a free surface. In a uniform field applied normal to the interface, the flat surface can lose stability and develop an ordered array of sharp peaks, the Rosensweig or normal-field instability \cite{CowleyRosensweig1967,Gailitis1977,BacriSalin1984,BoudouvisEtAlJMMM1987,GollwitzerEtAlJFM2007}. The peak state is a static interfacial pattern selected by the balance of gravity, capillarity, and field-induced stresses \cite{RosensweigBook,AbouWesfreidRoux2000}.

A ferrofluid surface can also undergo the familiar parametric (Faraday) instability of ordinary liquids: under periodic forcing, a flat interface destabilizes into standing waves \cite{BenjaminUrsell1954,CrossHohenberg1993}. Experiments combining vertical vibration with an applied magnetic field have used ferrofluids to explore how Faraday waves compete with the Rosensweig instability, suggesting that vibration can suppress peak formation and that tricritical-like behavior can emerge where the two onset lines intersect \cite{PetrelisFalconFauve2000,Muller1998PRE}. Ferrofluids thus provide a setting in which a static peak instability and a dynamical Faraday instability can coexist, motivating a basic question in nonequilibrium pattern selection: how do these instabilities organize and influence one another when the same interface is driven across their thresholds \cite{CrossHohenberg1993}?

\begin{figure}
\includegraphics[width=\columnwidth]{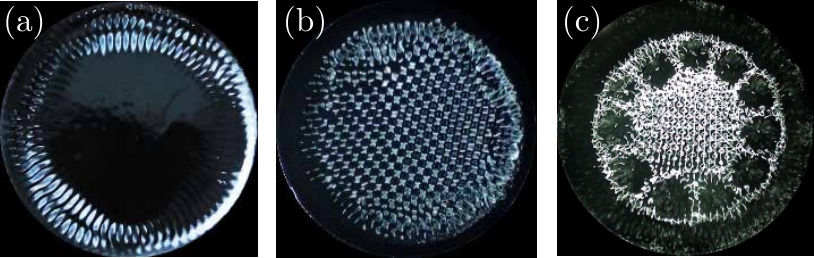}
\caption{Representative surface states under vertical sinusoidal magnetic driving $B(t)=B_0\sin(2\pi f t)$ with $f = 50 \, \mathrm{Hz}$. (a) Faraday-wave state showing boundary-influenced wavefronts ($B_0=14.0\,\mathrm{mT}$, $\eta=3.6\,\mathrm{cP}$). (b) Square-lattice Faraday state observed within the Faraday-wave regime ($B_0=14.0\,\mathrm{mT}$, $\eta=2.2\,\mathrm{cP}$). (c) Rosensweig-like peak state with localized peaks at larger drive ($B_0=16.0\,\mathrm{mT}$, $\eta=2.2\,\mathrm{cP}$). The container diameter is $60\,\mathrm{mm}$ and the liquid depth is $5\,\mathrm{mm}$.}
\label{fig:FF_image}
\end{figure}

A conceptually clean approach is to use a single drive that can access both instabilities, namely an oscillatory magnetic field applied normal to the interface. Oscillatory-field experiments have established that magnetic driving can generate standing-wave patterns with symmetries that differ from the familiar Rosensweig peak lattice \cite{BacriEtAlPRE1994,MahrRehberg1998,PiEtAlPRL2000}. Oscillatory fields have also been used to probe the nonlinear dynamics of Rosensweig peaks under modulation, including subharmonic responses and symmetry changes from triangular to square peak lattices \cite{BacriOrtonaSalin1991PRL,MahrRehbergPhysicaD1998,FriedrichsEngel2000,LangeLangerEngel2000}. In particular, Pi \textit{et al.} mapped a rich planform diagram under sinusoidal magnetic driving in the presence of a substantial DC bias \cite{PiEtAlPRL2000}. That work provides a key limiting case of oscillatory-field pattern formation and highlights the diversity of magnetically driven Faraday states. At the same time, the limited explored window and the DC bias complicate isolating a simple organizing picture that separates a Faraday-wave branch from a peak branch under oscillatory forcing.

Here we extend this program by scanning a substantially broader frequency range and reaching significantly larger AC amplitudes under strictly zero-mean driving. This enables a systematic phase diagram of ferrofluid surface states under a pure AC normal field that resolves two distinct and robust responses: a Faraday-wave regime that includes a stable square lattice and a Rosensweig-like peak--valley regime that is morphologically Rosensweig-like. The resulting phase boundaries are remarkably simple, with the Faraday-onset boundary well fit by a power law close to $\sqrt{f}$ and the peak boundary becoming essentially frequency independent at low viscosity. The square-lattice wave number $k_\ast=2\pi/a$ also grows linearly with frequency. These observations motivate a minimal surface-wave theory in which a single equation captures all observed features. We show below that this unified description explains why a zero-mean AC field can produce Rosensweig peaks, why the peak boundary is insensitive to frequency, why the wave boundary shifts systematically with frequency and viscosity, and why squares are selected over hexagons in the low-viscosity regime \cite{KumarTuckerman1994,ChenVinals1997,PiEtAlPRL2000}.

\begin{figure}
\includegraphics[width=\columnwidth]{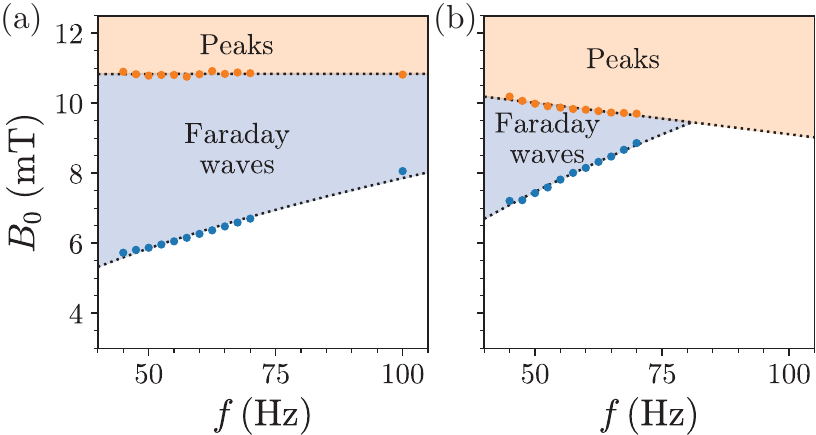}
\caption{Phase diagrams in the $(f,B_0)$ plane for two viscosities. Blue symbols indicate the Faraday instability onset and orange symbols indicate the Rosensweig-like peak onset as defined in the text. (a) Lower viscosity ($\eta=2.2\,\mathrm{cP}$, $\rho=0.87\,\mathrm{g/mL}$). (b) Higher viscosity ($\eta=3.6\,\mathrm{cP}$, $\rho=0.91\,\mathrm{g/mL}$). A stable square lattice appears within the Faraday-wave regime in an intermediate window between the two onsets. Dotted lines show fits (see main text for explanation).}
\label{fig:FF_PD}
\end{figure}

\PRLsec{Experimental phase diagram}
A commercial kerosene-based ferrofluid (HM01, Ferrofluidics) is placed in a cylindrical glass container of diameter $60\,\mathrm{mm}$ with a liquid fill height of $5\,\mathrm{mm}$. The container is centered inside a solenoidal coil (height $3.5\,\mathrm{cm}$, inner diameter $7.5\,\mathrm{cm}$, outer diameter $14\,\mathrm{cm}$) that produces a vertical magnetic field uniform to within $0.1\,\mathrm{mT}$ over the central region of the sample. The coil is driven sinusoidally so that the applied field is purely AC,
\begin{equation}
B(t)=B_0\sin(2\pi f t),
\end{equation}
where $f$ and $B_0$ denote the drive frequency and amplitude. The surface is imaged from above under oblique ring illumination. To vary dissipation we dilute the stock ferrofluid with petroleum ether, yielding a higher-viscosity sample with $\eta=3.6\,\mathrm{cP}$ and $\rho=0.91\,\mathrm{g/mL}$ and a lower-viscosity sample with $\eta=2.2\,\mathrm{cP}$ and $\rho=0.87\,\mathrm{g/mL}$. For each $f$, $B_0$ is increased in small steps, and after each step the system is allowed to relax until the observed state is reproducible.

Three regimes are observed as $B_0$ is increased at fixed $f$, as summarized in Fig.~\ref{fig:FF_image}. At low drive the interface is featureless within our imaging resolution. Above a first threshold, the surface develops standing-wave distortions, defining the Faraday-wave regime. The morphology of the Faraday-wave regime ranges from stripe-like standing waves influenced by the circular boundary (Fig.~\ref{fig:FF_image}(a)) to a remarkably regular square lattice (Fig.~\ref{fig:FF_image}(b)) as the drive strengthens. At sufficiently large drive, sharp localized peaks nucleate and the surface develops a dense peak--valley texture (Fig.~\ref{fig:FF_image}(c)), which we refer to as the Rosensweig-like peak state. Operationally, the Faraday-wave regime consists of low-amplitude surface-wave patterns, whereas the Rosensweig-like peak state is defined by the appearance of distinct peaks with heights of at least a few millimeters. The peak morphology is qualitatively distinct from the Faraday-wave regime and closely resembles the peak state of the Rosensweig instability known from static normal fields \cite{CowleyRosensweig1967,RosensweigBook,Gailitis1977,GollwitzerEtAlJFM2007}. Near the onset of the peak state, peaks typically appear on top of the preexisting wave background, so that wave order and peak nucleation can be simultaneously visible in real space.

Fig.~\ref{fig:FF_PD} compiles these observations into phase diagrams in the $(f,B_0)$ plane. For each $f$, the Faraday onset is defined as the smallest $B_0$ for which a standing-wave modulation is reproducibly observed in the central region. The Rosensweig-like peak onset is defined as the smallest $B_0$ for which localized peaks first appear. Motivated by the surface-wave theory developed in the next section, we use a power-law fit for the Faraday onset, which yields $B_{0,\mathrm{F}}(f)\approx (1.1\,\mathrm{mT})\,(f/1\,\mathrm{Hz})^{0.43}$
for the lower-viscosity sample and $B_{0,\mathrm{F}}(f)\approx (1.1\,\mathrm{mT})\,(f/1\,\mathrm{Hz})^{0.49}$
for the higher-viscosity sample. For the Rosensweig-like peak onset, the low-viscosity sample shows an essentially frequency-independent threshold,
$B_{0,\mathrm{R}}(f)\approx 10.83\,\mathrm{mT}$,
whereas the higher-viscosity sample is well described by
$B_{0,\mathrm{R}}(f)\approx (-0.020\,\mathrm{mT\,Hz^{-1}})\,f+10.90\,\mathrm{mT}$. Thus, within our experimental band, the Faraday onset is consistent with a power law close to $\sqrt{f}$, and the Rosensweig-like peak onset is weakly dependent on frequency and becomes effectively frequency independent at low viscosity. The square lattice is observed in an intermediate window between the two onsets, and lowering the viscosity broadens this window and stabilizes the lattice over a wider range of $(f,B_0)$.

These findings place earlier oscillatory-field results in a broader context. Pi \textit{et al.} \cite{PiEtAlPRL2000} demonstrated that sinusoidal magnetic driving can generate a wide variety of standing-wave planforms, including square and hexagonal lattices, under conditions with a substantial DC bias. The parameter regime explored in Ref.~\cite{PiEtAlPRL2000} corresponds to a subset of our Faraday-wave branch, in which the peak instability is not accessed and multiple wave planforms can compete. By scanning a wider frequency range and reaching larger strictly zero-mean AC amplitudes, we additionally resolve a sharply defined peak onset and obtain a phase diagram in which the response separates naturally into a Faraday-wave branch and a Rosensweig-like peak branch. The simple frequency dependence of the two onsets in Fig.~\ref{fig:FF_PD}---in particular, the near frequency independence of the Rosensweig-like peak onset at low viscosity---provides the empirical structure that guides the unified surface-wave description developed below.

\PRLsec{Surface-wave theory}
The phase diagram can be organized by separating a static softening mechanism from a dynamical Floquet mechanism, both generated by the same oscillatory magnetic drive. We describe the interface by its height field $\zeta(\mathbf{r},t)$ and expand in lateral Fourier modes $\zeta_{\mathbf{k}}(t)$ with $k=|\mathbf{k}|$. Linearizing the free-surface dynamics gives, for each mode, a damped oscillator with a field-dependent restoring term \cite{LandauLifshitzFM,ZelazoMelcher1969,RosensweigBook},
\begin{equation}
\ddot{\zeta}_{\mathbf{k}}+2\Gamma(k)\dot{\zeta}_{\mathbf{k}}+\Omega^2(k;H)\,\zeta_{\mathbf{k}}=0,
\label{eq:eom}
\end{equation}
where $\Gamma(k)$ encodes viscous dissipation and $\Omega(k;H)$ is the surface-wave frequency in a normal magnetic field. In the deep-layer limit, the leading structure of the dispersion can be written as \cite{CowleyRosensweig1967,ZelazoMelcher1969,RosensweigBook,BrowaeysEPJB1999,BrowaeysBJP2001}
\begin{equation}
\Omega^2(k;H)=gk+\frac{\sigma}{\rho}k^3-\alpha\,H^2 k^2,
\qquad
\alpha=\frac{\mu_0}{\rho}\,\mathcal{S}(\chi),
\label{eq:disp}
\end{equation}
where $g$ is the gravitational acceleration, $\sigma$ is the surface tension, $\rho$ is the density, $\mu_0$ is the vacuum permeability, and $\mathcal{S}(\chi)$ is a dimensionless factor set by magnetic boundary conditions and the (effective) susceptibility $\chi$. For most part of this work, we have neglected finite layer depth correction. The essential point is that the magnetic contribution enters as $H^2$, reflecting the quadratic Maxwell stress at the interface \cite{RosensweigBook,BrowaeysBJP2001,ZelazoMelcher1969}.

For the present low-frequency driving we adopt a magnetoquasistatic approximation and treat the field-dependent restoring term as instantaneous $\Omega^2(k ; H(t))$. Consider a purely AC drive $H(t)=H_0\sin(\omega t)$, one has
\begin{equation}
H^2(t)=H_{\mathrm{rms}}^2\Bigl(1-\cos 2\omega t\Bigr),
\qquad
H_{\mathrm{rms}}=\frac{H_0}{\sqrt{2}}.
\label{eq:H2}
\end{equation}
Substituting Eq.~(\ref{eq:H2}) into Eqs.~(\ref{eq:eom}) and (\ref{eq:disp}) yields a damped Mathieu equation \cite{MilesHenderson1990,KumarTuckerman1994},
\begin{equation}
\ddot{\zeta}_{\mathbf{k}}+2\Gamma(k)\dot{\zeta}_{\mathbf{k}}
+\Bigl[\Omega^2(k;H_{\mathrm{rms}})-\epsilon(k)\cos(2\omega t)\Bigr]\zeta_{\mathbf{k}}=0,
\label{eq:mathieu}
\end{equation}
where $\epsilon(k)=\alpha H_{\mathrm{rms}}^2 k^2$. Eq.~(\ref{eq:mathieu}) contains two physically distinct effects from the same drive. The time-averaged component $H_{\mathrm{rms}}^2$ produces a static softening and can trigger a Rosensweig-type instability when $\Omega^2(k;H_{\mathrm{rms}})$ becomes negative for some $k$ \cite{CowleyRosensweig1967,RosensweigBook,BacriSalin1984}. The onset is determined by $\Omega^2(k_c;H_c)=0$ and $\partial_k \Omega^2(k;H_c)|_{k_c}=0$, which for Eq.~(\ref{eq:disp}) give \cite{CowleyRosensweig1967,RosensweigBook}
\begin{equation}
k_c=\sqrt{\frac{\rho g}{\sigma}},
\qquad
\mu_0 H_c^2=\frac{2\sqrt{\rho g\sigma}}{\mathcal{S}(\chi)}.
\label{eq:rosen}
\end{equation}
Under a pure AC drive the criterion is $H_{\mathrm{rms}}\gtrsim H_c$, so the Rosensweig-like peak onset in the experiment is governed by the RMS amplitude and is insensitive to the drive frequency to leading order. This directly accounts for the nearly frequency-independent Rosensweig-like peak onset in Fig.~\ref{fig:FF_PD}(b).

The oscillatory term in Eq.~(\ref{eq:mathieu}) produces a Faraday instability. Because the magnetic stress is quadratic in $H$, a sinusoidal drive modulates the coefficient at $2\omega$, and the principal instability tongue is subharmonic centered at $\omega$ \cite{BenjaminUrsell1954,MilesHenderson1990,KumarTuckerman1994}. In the weak-damping limit the onset of this tongue satisfies $\epsilon_*(k_\ast)\simeq 4\,\Gamma(k_\ast)\,\omega$ \cite{BenjaminUrsell1954,MilesHenderson1990,KumarTuckerman1994}. With $\epsilon(k)=\alpha H_{\mathrm{rms}}^2 k^2$ this gives
\begin{equation}
H_{\mathrm{rms},*}^2 \simeq \frac{4\,\Gamma(k_\ast)}{\alpha\,k_\ast^{2}}\,\omega.
\label{eq:Hrms_threshold}
\end{equation}
In the deep-layer limit $\Gamma(k)\approx 2\nu k^2$ with $\nu=\eta/\rho$ \cite{LandauLifshitzFM}, Eq.~(\ref{eq:Hrms_threshold}) reduces to a simple $k_*$ independent form,
\begin{equation}
    H_{\mathrm{rms},*}\simeq \sqrt{\frac{8\nu}{\alpha}\,\omega},
\end{equation}
i.e., a wave-onset threshold linear in the RMS magnetic pressure and decreasing with viscosity. Since $B_0\propto H_0=\sqrt{2}\,H_{\mathrm{rms}}$, this scaling implies $B_{0,*}\propto \sqrt{f}$. This is consistent with our power-law fits of the Faraday onset in Fig.~\ref{fig:FF_PD}, which yield exponents $0.49$ and $0.43$.

In the same regime, magnetic softening drives the surface-wave dispersion toward a shallow finite-$k$ minimum near the Rosensweig point. A local expansion about this minimum yields an approximately conical form, $\Omega(k)\approx c\,|k-k_c|$, over the window where the curvature is small \cite{CowleyRosensweig1967,RosensweigBook}. This magnetically softened dispersion has been characterized directly in ferrofluids through dispersion measurements and wave-drag (wave-resistance) experiments under vertical fields \cite{BrowaeysEPJB1999,BrowaeysBJP2001}. The conical form implies a linear wavevector--frequency selection: the dominant Faraday mode satisfies $\Omega(k_\ast)\approx \omega$, giving \cite{CowleyRosensweig1967,RosensweigBook}
\begin{equation}
k_\ast(f)\simeq k_c+\frac{\omega}{c}.
\label{eq:qlinear}
\end{equation}
In the deep-layer limit of Eq.~(\ref{eq:disp}), the conical slope at the Rosensweig point is fixed by capillary and gravity scales \cite{CowleyRosensweig1967,RosensweigBook},
\begin{equation}
c=\left(\frac{g\sigma}{\rho}\right)^{1/4}.
\label{eq:c_sigma}
\end{equation}

The competition between square versus hexagonal lattices is constrained by the time-translation symmetry of the parametric coefficient $H^2(t)$ in Eq.~(\ref{eq:mathieu}) \cite{BenjaminUrsell1954,KumarTuckerman1994,ChenVinals1997,CrawfordKnobloch1991}. For a zero-mean drive $H(t)$ at frequency $\omega$, $H^2(t)$ has periodicity $T_p=\pi/\omega$ and the linear problem is invariant under $t\to t+T_p$. At onset this yields Floquet classes
\begin{equation}
\zeta(t+T_p)=\pm \zeta(t),
\end{equation}
where the $+$ class is the harmonic response and the $-$ class is the subharmonic response with respect to the coefficient. In our parameter range the onset is observed in the subharmonic class. Writing $\zeta\sim\sum_j(A_j e^{i\mathbf k_j\cdot\mathbf r}+\mathrm{c.c.})$, the subharmonic condition enforces a sign-flip symmetry $A_j\to -A_j$, which forbids quadratic terms in the standing-wave amplitude equations and thus eliminates the leading-order resonant triad mechanism for hexagons. With this quadratic mechanism absent, cubic competition near threshold generically stabilizes two-mode states, hence squares \cite{ChenVinals1997,CrawfordKnobloch1991,PiEtAlPRL2000}.

A DC bias removes precisely this symmetry. For $H(t)=H_0+\Delta H\sin(\omega t)$,
\begin{equation}
H^2(t)=H_0^2+2H_0\Delta H\sin(\omega t)+\frac{\Delta H^2}{2}\Bigl(1-\cos 2\omega t\Bigr),
\label{eq:H2dc}
\end{equation}
the $\omega$ term implies $H^2(t+T_p)\neq H^2(t)$, so the subharmonic Floquet class is no longer symmetry-protected and quadratic triad interactions become symmetry-allowed \cite{ChenVinals1997,PiEtAlPRL2000}. Consistent with this symmetry lifting, we experimentally observe a periodicity doubling from subharmonic response at $\omega$ to harmonic response at $2 \omega$ once an additional DC magnetic field is applied.

\begin{figure}
\includegraphics[width=\columnwidth]{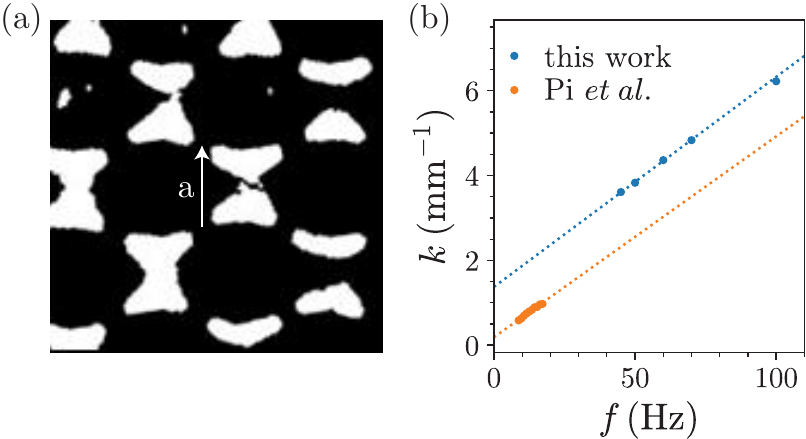}
\caption{Wave number of the square-lattice state. (a) Real-space definition of the lattice spacing $a$ and the wave number $k\equiv 2\pi/a$. (b) Measured $k$ versus drive frequency $f$ for the low-viscosity sample ($\rho=0.87\,\mathrm{g/mL}$, $\eta=2.2\,\mathrm{cP}$) within the square-lattice window of the Faraday-wave regime. The data are well fit by a linear relation $k(f)=k_0+\omega/c_{\mathrm{eff}}$. Orange symbols show the square-branch data of Pi \textit{et al.} \cite{PiEtAlPRL2000} replotted after converting their wave-number convention by a factor of $2\pi$ to match $k=2\pi/a$.}
\label{fig:FF_period}
\end{figure}

\PRLsec{Wavevector}
The square-lattice Faraday state provides a direct probe of wavevector selection. From each top-view image we extract the lattice spacing $a$ and define the wave number $k\equiv 2\pi/a$, as indicated in Fig.~\ref{fig:FF_period}(a). Fig.~\ref{fig:FF_period}(b) shows that $k$ grows linearly with frequency over the full accessible band. Writing the fit in the form suggested by Eq.~(\ref{eq:qlinear}),
\begin{equation}
k(f)=k_0+\frac{\omega}{c_{\mathrm{eff}}},
\qquad
\omega=2\pi f,
\label{eq:qfit}
\end{equation}
we obtain $c_{\mathrm{eff}}=0.13\,\mathrm{m/s}$ and $k_0=1.38\,\mathrm{mm^{-1}}$. In the ideal deep-layer limit and asymptotically close to the Rosensweig point one expects $k_0\to k_c=\sqrt{\rho g/\sigma}$ and $c_{\mathrm{eff}}\to c=(g\sigma/\rho)^{1/4}$ as in Eqs.~(\ref{eq:rosen}) and (\ref{eq:c_sigma}), whereas finite-depth effects, field-dependent magnetic response, and finite-band linearization can shift the apparent offset away from $k_c$ \cite{BrowaeysEPJB1999,BrowaeysBJP2001,Muller1998PRE}.

A useful comparison is the square-lattice branch reported by Pi \textit{et al.} \cite{PiEtAlPRL2000}. We note that the wave-number convention in Ref.~\cite{PiEtAlPRL2000} differs by a factor of $2\pi$ from the convention $k=2\pi/a$ used here, and we rescale accordingly when plotting their data in Fig.~\ref{fig:FF_period}(b). With this conversion, the slope of their square branch matches ours within their stated uncertainty, indicating a comparable $c_{\mathrm{eff}}$ despite differences in working fluid and driving protocol. In contrast, the offset differs markedly. Using the deep-layer relations in Eqs.~(\ref{eq:rosen}) and (\ref{eq:c_sigma}), one has $k_c=\sqrt{\rho g/\sigma}=g/c^{2}$, so the measured $c_{\mathrm{eff}}$ implies $k_c\approx 560\,\mathrm{m^{-1}}$, consistent with the separation between Rosensweig-like peaks at large $B_0$ in our experiments. Meanwhile, the fitted intercepts correspond to $k_0\approx 189\,\mathrm{m^{-1}}$ for Ref.~\cite{PiEtAlPRL2000} and $k_0\approx 1.38\times 10^{3}\,\mathrm{m^{-1}}$ for our data, bracketing the scale inferred from $c_{\mathrm{eff}}$. We attribute this discrepancy in the apparent offset to finite-depth effects \cite{Muller1998PRE}.

\PRLsec{Discussion}
By mapping the full $(f,B_0)$ phase diagram under zero-mean oscillatory magnetic driving, we show that a single AC field can organize ferrofluid surface dynamics into a static peak branch and a dynamical standing-wave branch with sharply separated onsets. The simplicity of the observed scalings, including a power-law Faraday onset close to $\sqrt{f}$, a weakly frequency-dependent Rosensweig-like peak onset, and a linear $k_\ast(f)$, enables a minimal surface-wave description in which RMS softening and Floquet forcing arise from the same quadratic magnetic stress \cite{CowleyRosensweig1967,BrowaeysBJP2001,KumarTuckerman1994}. In this perspective, earlier DC-biased oscillatory-field studies \cite{PiEtAlPRL2000} appear as a controlled limit of the broader driven problem. More generally, strong AC magnetic driving offers a compact route to explore how static and Floquet instabilities meet and compete on a single interface, with viscosity and bias providing independent handles on temporal symmetry and planform selection \cite{CrossHohenberg1993}.

\textbf{Acknowledgments.}
We thank Prof.\ Luwei Zhou and Prof.\ Jiping Huang (Fudan University) for valuable discussions throughout this project. We also thank Hanjue Zhu (now at the Department of Astronomy and Astrophysics, The University of Chicago) for assistance with manuscript preparation. The experiments were carried out at No.\ 2 High School of East China Normal University and were supported by Fudan University.

\bibliography{main}

\end{document}